\documentclass[lettersize,journal]{IEEEtran}
\usepackage{amsmath,amsfonts}
\usepackage{algorithm}
\usepackage{algpseudocode}
\usepackage{array}
\usepackage[caption=false,font=normalsize,labelfont=sf,textfont=sf]{subfig}
\usepackage{textcomp}
\usepackage{stfloats}
\usepackage{url}
\usepackage{verbatim}
\usepackage{graphicx}
\usepackage{cite}
\usepackage{multirow}

\usepackage[export]{adjustbox}

\algblock{Input}{EndInput}
\algnotext{EndInput}
\algblock{Output}{EndOutput}
\algnotext{EndOutput}
\newcommand{\Desc}[2]{\State \makebox[6em][l]{#1}#2}

\begin{document}

\title{Real-time Event Recognition of Long-distance Distributed Vibration Sensing with Knowledge Distillation and Hardware Acceleration}

\author{Zhongyao Luo, Hao Wu, Zhao Ge, and Ming Tang,~\IEEEmembership{Senior Member,~IEEE,}
\thanks{This work was supported in part by the National Key Research and Development Program of China under Grant 2021YFB2800902; in part by the National Natural Science Foundation of China under Grant 62225110; and in part by the innovation Fund of WNLO. (Corresponding author: Hao Wu).}
\thanks{The authors are with the Wuhan National Laboratory for Optoelectronics, Next Generation Internet Access National Engineering Laboratory, and Hubei Optics Valley Laboratory, School of Optical and Electronic Information, Huazhong University of Science and Technology, Wuhan 430074, China (email: zluo@hust.edu.cn, wuhaoboom@hust.edu.cn, d202280977@hust.edu.cn, tangming@mail.hust.edu.cn)}}

\markboth{\textit{T\MakeLowercase{his work has been submitted to the} IEEE \MakeLowercase{for possible publication.} C\MakeLowercase{opyright may be transferred without notice, after which this version may no longer be accessible.}}%
}{}

\maketitle

\begin{abstract}
Fiber-optic sensing, especially distributed optical fiber vibration (DVS) sensing, is gaining importance in internet of things (IoT) applications, such as industrial safety monitoring and intrusion detection. 
Despite their wide application, existing post-processing methods that rely on deep learning models for event recognition in DVS systems face challenges with real-time processing of large sample data volumes, particularly in long-distance applications. 
To address this issue, we propose to use a four-layer convolutional neural network (CNN) with ResNet as the teacher model for knowledge distillation. This results in a significant improvement in accuracy, from 83.41\% to 95.39\%, on data from previously untrained environments. 
Additionally, we propose a novel hardware design based on field-programmable gate arrays (FPGA) to further accelerate model inference. This design replaces multiplication with binary shift operations and quantizes model weights, enabling high parallelism and low latency.
Our implementation achieves an inference time of 0.083 ms for a spatial-temporal sample covering a 12.5 m fiber length and 0.256 s time frame. 
This performance enables real-time signal processing over approximately 38.55 km of fiber, about $2.14\times$ the capability of an Nvidia GTX 4090 GPU. The proposed method greatly enhances the efficiency of vibration pattern recognition, promoting the use of DVS as a smart IoT system. The data and code are available at https://github.com/HUST-IOF/Efficient-DVS.
\end{abstract}


\section{Introduction}
\begin{figure}[!t]
  \centering
  \includegraphics[width=\linewidth]{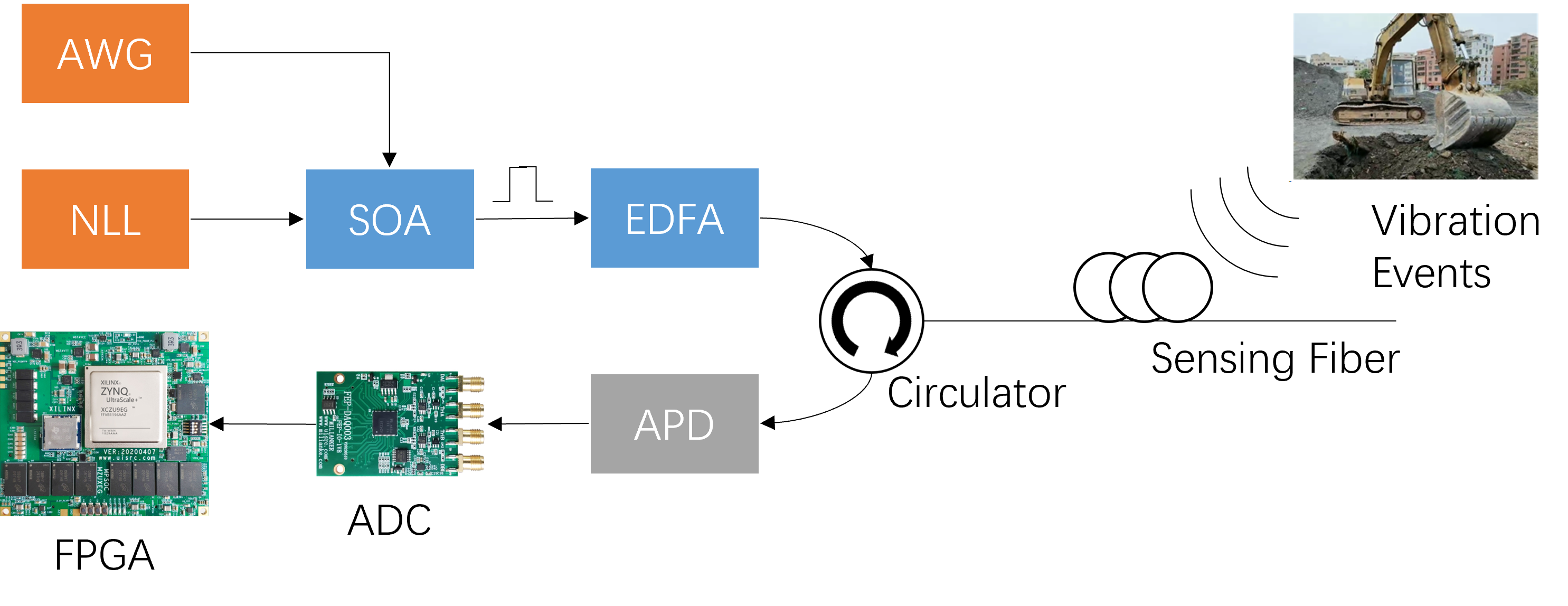}
  \caption{Schematic of DVS system. NLL: narrow linewidth laser, SOA: semiconductor optical amplifier, AWG: arbitrary waveform generator, EDFA: erbium-doped fiber amplifier, FUT: fiber under test, APD: avalanche photodetector, DAQ: data acquisition card.}
  \label{DVS_SYS}
\end{figure}

\IEEEPARstart{D}{istributed} optical fiber vibration sensing (DVS) technology, based on phase-sensitive optical time-domain reflectometry (phi-OTDR), uses optical fibers as sensing elements to provide dense, continuous measurements of vibration and recognize various intrusion events. This technology has been applied as smart internet of things (IoT) systems in various areas in recent years, including industrial safety monitoring and intrusion detection. Recent applications of DVS technology include railway safety monitoring \cite{yangRailwayIntrusionEvents2022, yangRealTimeFOTDRVibration2022a}, perimeter security \cite{lyuDistributedOpticalFiber2020, Huang2020AnER}, pipeline monitoring \cite{yangEarlySafetyWarnings, zhuDistributedOpticalFiber2023}, and smart cities \cite{wuSmartFiberOpticDistributed2023}.
As shown in Fig. \ref{DVS_SYS}, the phi-OTDR system captures Rayleigh backscattering light across the fiber, with phase variations induced by external vibrations providing valuable event information.

Despite its capabilities, DVS technology faces several challenges in IoT applications. The primary issue is the variability of signals due to complex and changing environments and diverse intrusion types, which limits event recognition accuracy. 
Traditional methods often rely on hand-crafted algorithms for data preparation and feature extraction, followed by classic machine learning for recognition\cite{xuPatternRecognitionBased2017, wuFeatureExtractionIdentification2017, wuDynamicTimeSequence2019a, caoPracticalPatternRecognition2015, wangEventIdentificationBased2019, jiaNearestNeighborAlgorithmBased2019}. 
These approaches can be time-consuming and complex, requiring extensive expertise in signal processing.
To solve these problems, deep learning models (DLMs) have been introduced to the field. Initially, DLMs were introduced for pattern recognition based on extracted features \cite{wuFeatureExtractionIdentification2017}. 
With the development of more sophisticated and computationally powerful DLMs, they are capable of automatically extracting important features from noisy data and simultaneously handling feature extraction and pattern recognition tasks. 
The introduction of complex DLMs brings about improved performance and reduces the design workload. Consequently, they have received more attention in recent years. Table \ref{dvs_list} shows several typical methods of this type of application, where $GAF$ represents gramian angular field, $S-T$ denotes space-time, and $F-T$ is frequency-time.
\begin{table}[h]
\centering
\caption{Previous Work on DLM-based DVS Algorithms and Our Work}
\label{dvs_list}
\resizebox{\linewidth}{!}{
\begin{tabular}{|c|c|c|c|c|}
\hline
Model & Input Form & Parameter Number & Accuracy (\%) & Pub. Year\\
\hline
VGG16 & GAF Figure & 138M & 97.67 & 2020\cite{lyuDistributedOpticalFiber2020}\\
\hline
YoloV3 & S-T Image & 45.5M & - & 2021\cite{shaPIGTrackingUtilizing2021a}\\
\hline
Yolo-A30 & S-T Image & 65M & 99.2 & 2022\cite{yangRealTimeFOTDRVibration2022a} \\
\hline
Yolo & S-T Image & 61M & 96.1 & 2022\cite{xuRealTimeMultiClassDisturbance2022b} \\
\hline
MTL & S-T Image & 1.13M & 99.46 & 2023\cite{wuSmartFiberOpticDistributed2023}\\
\hline
Swin-T & GAF Figure & 28M & 92.44 & 2023\cite{zhuDistributedOpticalFiber2023}\\
\hline
Resnet-152 & F-T Image & 60M & 96.67 & 2023\cite{jinPatternRecognitionDistributed2023}\\
\hline
CNN-4 (This work) & S-T Image & 30K & 95.39 & -\\
\hline
\end{tabular}
}
\end{table}

Another challenge in applying DVS technology in IoT applications is the large volume of captured data, which scales with the length of the fiber. Consequently, monitoring long-distance fibers requires processing vast amounts of data. Moreover, applications such as pipeline monitoring and earthquake detection demand real-time or near-real-time responses from smart IoT systems based on DVS technology to minimize potential property damage.
In addition, the use of high-capacity DLMs significantly increases the demand for computational resources. In real-world scenarios, where resources are often limited, achieving real-time event recognition is typically feasible only for short-distance fiber monitoring \cite{Xu2022RealTimeMD}.
To improve the computational efficiency of the DVS system, there are typically three solutions. 
The first is data selection or down-sampling, which involves using prior knowledge and data analysis to identify and retain valuable signal sections or sample points\cite{yangRealTimeFOTDRVibration2022a, wuSmartFiberOpticDistributed2023}. While this reduces the input data volume, it can lead to false negatives due to its reliance on hand-crafted algorithms.
The second solution is model compression, which seeks to lower computational complexity and resource demands. However, existing approaches often require models with at least millions of parameters to maintain performance and generalizability\cite{wuSmartFiberOpticDistributed2023},  leaving room for further optimization to develop smaller and more efficient models. 
The third solution is the use of high-performance hardware, such as graphics processing units (GPUs) and cloud computing. While this approach can significantly increase computational power, it is often expensive and demands high power consumption, as well as a reliable communication system to handle the large volume of sample data.

This paper proposes enhancing the signal processing capabilities of DVS systems through the use of shallow convolutional neural network (CNN) coupled with knowledge distillation and hardware acceleration. We introduce a shallow CNN as a more efficient alternative to DLMs, reducing model complexity and computational resource requirements. we identify the generalizability limitation of low-capacity models and propose to use the knowledge distillation technique as a solution. 
To further improve the  data processing throughput, we design an efficient hardware acceleration scheme based on a field programmable gate array (FPGA). An analog-to-digital Converter (ADC) is integrated with the FPGA to minimize latency during data transfer, as depicted in Fig. 1. 
The proposed scheme leverages FPGA’s flexibility to optimize model performance by replacing multiplication operations with shift operations, which are less resource-intensive. This shift-add structure, as we refer to it, allows for greater parallelism and faster computation, constrained only by the FPGA's logic resources rather than its limited digital signal processing (DSP) resources\cite{Meng2021FixyFPGAEF} on the chip.
Additionally, we propose a shift parameter quantization technique, where model weights are quantized into integers corresponding to the number of binary shift operations. This approach enables real-time processing of the substantial data volumes captured by the DVS system on a mainstream industrial FPGA.

In a word, the contribution of this work can be listed as follows:
\begin{itemize}
    \item We propose a novel scheme for designing and implementing real-time DVS algorithms and evaluate it using real-world DVS datasets. The results demonstrate comparable performance and generalizability to deep learning models, while achieving real-time recognition over long-distance fibers. 

    \item We propose to use shallow CNN models, coupled with knowledge distillation, to develop a lightweight model that offers high performance and generalizability. This approach effectively addresses the throughput limitation brought by the high computational complexity of existing algorithms. 

    \item We design a novel hardware acceleration scheme for shallow CNNs, utilizing shift-add structures and shift parameter quantization to significantly improve efficiency.

\end{itemize}

\section{Method}
\subsection{Knowledge Distillation based on logits}
\begin{figure*}[!t]
  \centering
  \includegraphics[width=\textwidth]{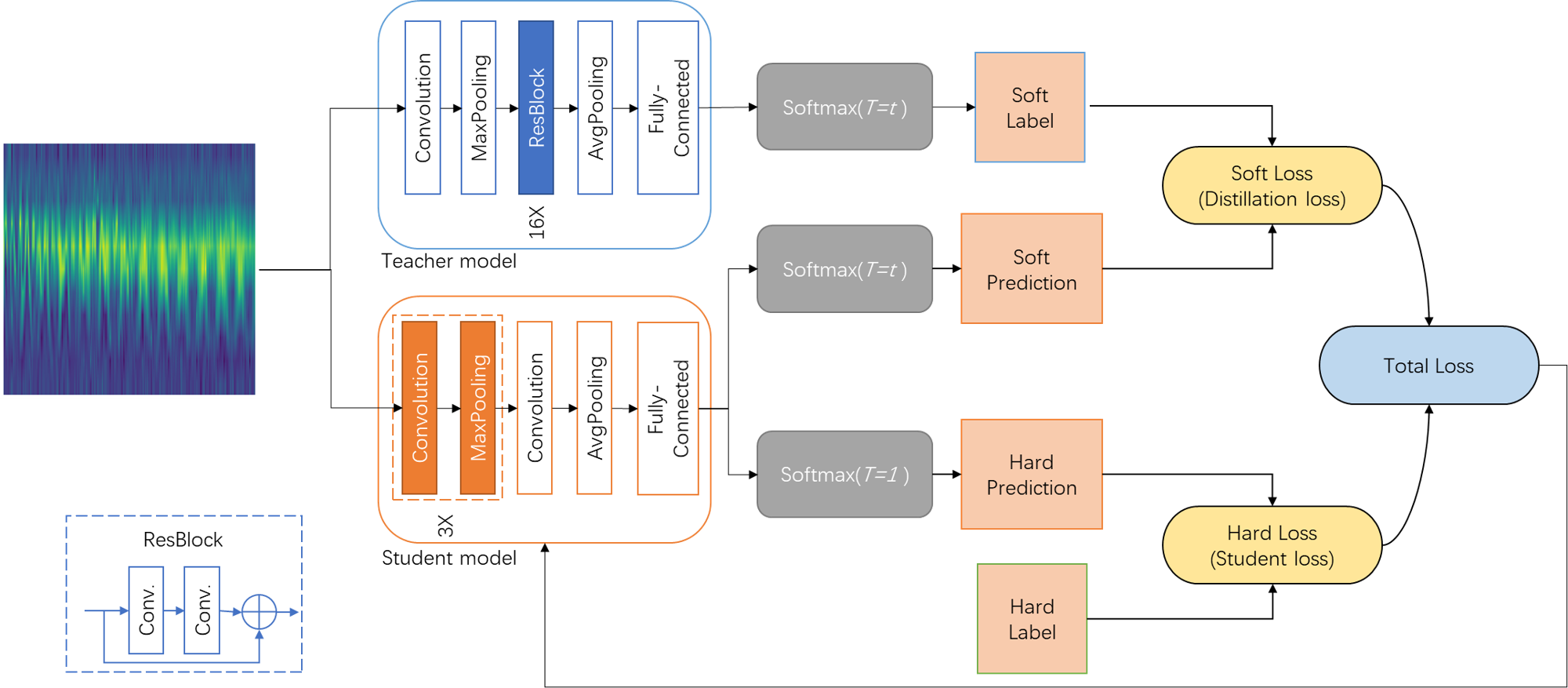}
  \caption{Schematic of KD process.}
  \label{KD_PROC}
\end{figure*}
It is generally believed that large models, that is, models with more complex and efficient structural design and huge number of parameters, tend to be able to extract higher-level features, so that it can achieve better results in various tasks. In other words, large models often have better generalizability, and can better handle signals with various noise. On the other hand, the model inference process can be considered as finishing the specified task based on the ’knowledge’ contained in model structure and its parameters. 
The idea of knowledge distillation is to extract the knowledge from large models (teacher models), and transfer the knowledge to small models (student model), so that the performance and generalizability of the later one can be improved. In the knowledge distillation method based on logits\cite{hintonDistillingKnowledgeNeural2015}, the output prediction distribution of the model is considered to contain valuable knowledge. This distribution represents the probabilities of classifying the input sample into each class.
The distribution can be obtained through introducing a temperature parameter $T$ to the softmax function at output layer. The function can be expressed as:
\begin{equation}
\text{Softmax}(\mathbf{x}/T)_i = \frac{e^{x_i / T}}{\sum_j e^{x_j / T}}
\end{equation}
where \( \mathbf{x} = [x_1, x_2, \ldots, x_n] \) is the input vector, which is the output of the network in this case.
The knowledge is then transferred by guiding the student model to imitate the behavior of the teacher model.
The process is to minimizing a loss function that measures the difference between the outputs of the teacher network and the student network, and also the difference between the outputs of the student network and the correct results, as shown in Fig. \ref{KD_PROC}. 
The loss function usually consists of a standard classification loss, like cross-entropy, and a distillation loss that measures the difference between the softened probabilities generated by the teacher and student networks. Formally, the loss function \( \mathcal{L} \) can be expressed as:
\begin{equation}
\begin{split}
\mathcal{L} & = \alpha \cdot \text{CE}(\mathbf{y}_s, \mathbf{y}_{\text{true}}) \\
& \quad + (1 - \alpha) \cdot \text{KL}(\text{Softmax}(\mathbf{y}_t/T), \text{Softmax}(\mathbf{y}_s/T))
\end{split}
\end{equation}
where \( \text{CE} \) denotes the cross-entropy loss, \( \text{KL} \) represents the Kullback-Leibler divergence, \( \mathbf{y}_{\text{true}} \) is the true label, $\alpha$ represents wegiht between two kind of losses, \( \mathbf{y}_t \) and \( \mathbf{y}_s \) denote the outputs of the teacher and student networks respectively. 
Through this optimization process, knowledge distillation enables the creation of compact yet powerful neural networks capable of retaining the performance of their larger counterparts.

\subsection{FPGA-Based Hardware Acceleration}
\subsubsection{Parallel Architecture}
\begin{figure}[h]
  \centering
  \includegraphics[width=\linewidth]{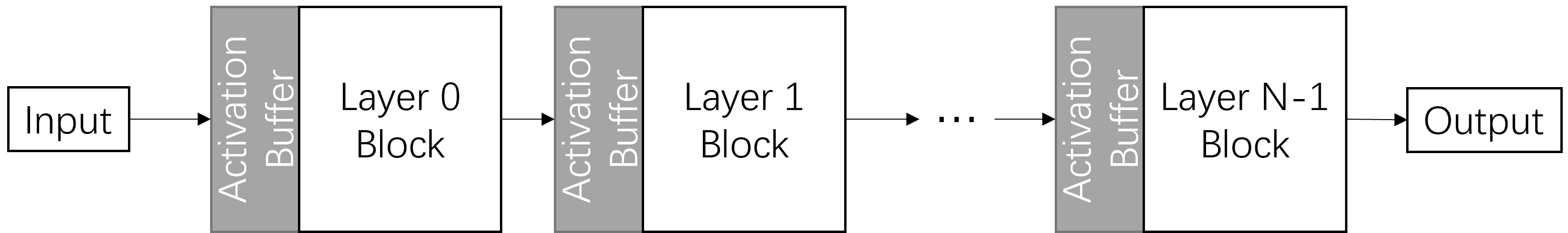}
  \caption{Schematic of pipelined structure.}
  \label{pipeline}
\end{figure}

The main problem of implementing CNN on FPGA is the storage of intermediate data generated during the calculation process and the parameters of the model\cite{Kang2022AoCStreamAC}.
There are two typical solutions to the problem.
One option is to use external memory to expand the storage capacity\cite{Aarrestad2021FastCN, Anupreetham2021EndtoEndFO, Pang2020AnEI,Meng2021FixyFPGAEF}.
This type of design usually combines on-chip storage and external storage to form a cache system.
The on-chip computing unit and memory access pattern is designed in accordance with the bandwidth limitation of the cache system.
The model is divided into multiple computing tasks suitable for the computing unit. As the tasks are completed in sequence, the calculation of the entire model is also completed.
Another option is to compress the model and store all intermediate data and parameters on-chip\cite{Kang2022AoCStreamAC, Meng2021FixyFPGAEF}.
Such design typically adopts a pipeline structure, implementing the model as multiple parallel modules to improve the throughput of the design.
In this paper, in order to achieve faster inference efficiency, the later one is adopted.
Each layer of the model is implemented as a separate module as shown in Fig. \ref{pipeline}.

\subsubsection{Activation Buffering} 
The activation buffer, as shown in Fig. \ref{pipeline}, stores intermediate data from the preceding layer. 
A typical convolution and max pooling layer can be expressed as follows:

\begin{equation}
\begin{split}
    O_{conv}[m, x, y] & = \sum_{n=0}^{N-1} \sum_{p=0}^{P-1} \sum_{q=0}^{Q-1} I[n, x + p, y + q] \\
    & \quad * K[m, n, p, q] + B[m] \\
\end{split}
\end{equation}

\begin{equation}
        O_{max}[n, x, y] = \max\{I[n, x : x + P, y : y + Q]\}
\end{equation}
Here, the symbols \(O\), \(I\), \(M\), \(N\), \(P\), \(Q\), \(K\), and \(B\) represent the following components: \(O\) denotes the output feature map, \(I\) represents the input feature map, \(M\) corresponds to the number of output channels, \(N\) denotes the number of input channels, \(P\) represents the kernel height, \(Q\) represents the kernel width, \(K\) symbolizes the convolution kernel, and \(B\) signifies the bias term.

\begin{figure}[h]
  \centering
  \includegraphics[width=2.5in]{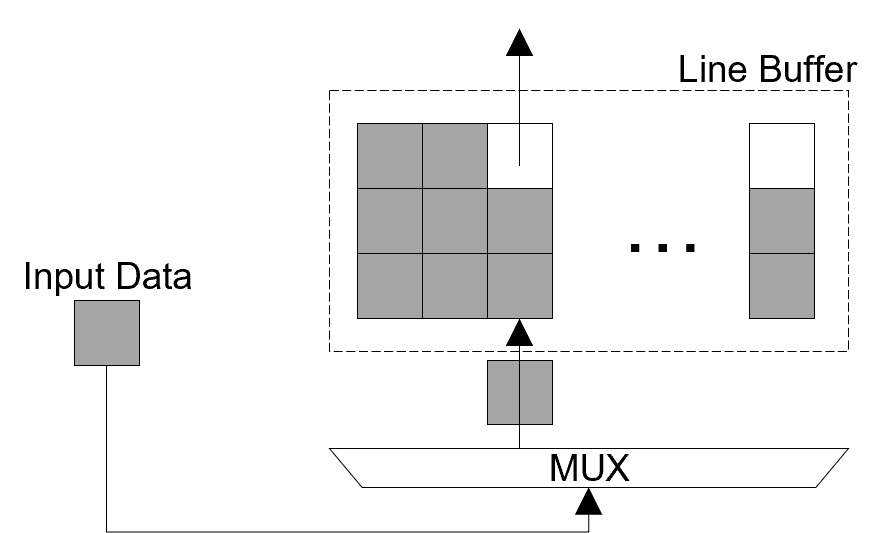}
  \caption{Line buffer structure.}
  \label{line_buffer}
\end{figure}

Obviously, both convolution layers and pooling layers only require the input feature map data within the selected window.
Assume the input is a spatial-temporal figure, where rows represent discrete time points and columns represent spatial positions. The data is processed in row-major order.
Then the activation buffer only needs to store \(P\) rows from the input feature map during the computation process.
Thus, the buffer mechanism is designed as shown in Fig. \ref{line_buffer}. 

When new data is inputted, the corresponding column shifts the existing data up by one unit, and the new data is filled into the bottom unit. 
This design allows the buffer to store just \( P-1 \) rows and \( Q \) data from the last row to start the convolution operation.
Thus, the design only requires  \(P\) rows of storage capacity. 
Moreover, reading new data and calculation can be performed in parallel to achieve lower latency.
The convolution layer only needs to buffer $(P-1-S)*W + Q$ data for each channel instead of the complete input feature map to start operation, where \(W\)and \(S\) denote the width of input feature map and the stride respectively.

\subsubsection{Shift-Add Structure}
\begin{figure}[h]
  \centering
  \includegraphics[width=\linewidth]{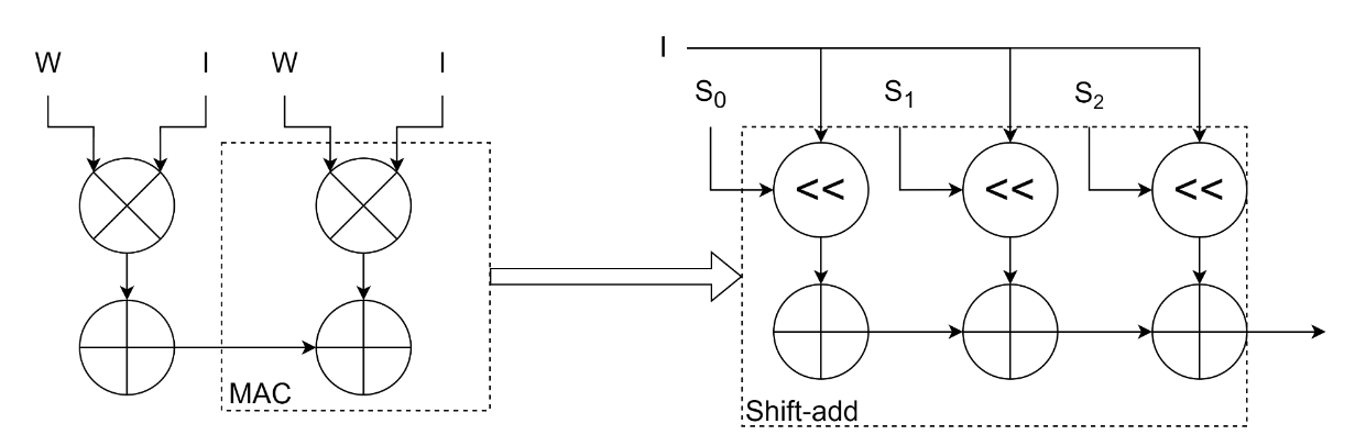}
  \caption{Shift-add and MAC structure.}
  \label{sa}
\end{figure}

The main operation of convolutional neural networks is multiplication and accumulation (MAC).
DSP block is required to implement MAC units on FPGA.
The limited number of DSP blocks restricts the number of MAC units that can be implemented on-chip. 
As a result, the throughput of the design is restricted.
To avoid this restriction, it is proposed to use the shift and accumulation units that do not require DSP resources for implementation to replace the MAC units, as shown in Fig. \ref{sa}. Digital values can be represented as a sum of powers of 2, as illustrated in Equation \ref{quan_1} with exponents denoted as \(s_0, s_1 , \ldots, s_k\). Multiplication with power of two is essentially performing shift operations on multiplier. This allows a MAC operations to be losslessly replaced by a series of shift and accumulation operations, as shown in Equation \ref{quan_2}. 

\begin{equation}
    W= 2^{s_0}+ 2^{s_1}+ \cdots +2^{s_k}
    \label{quan_1}
\end{equation}

\begin{equation}
\begin{split}
     I \cdot W &= I \cdot (2^{s_0}+ 2^{s_1}+ \cdots +2^{s_k})\\
            &= I << s_0 + I << s_1 + \cdots + I << s_k   
\end{split}
\label{quan_2}
\end{equation}

\subsubsection{Shift Parameter Quantization}
After converting all the MAC operations to shift and accumulation operations, the powers of two in the above equation \ref{quan_1} and \ref{quan_2} can be seen as separate parameters representing number of shift and accumulation operations, which are referred as shift parameters in this paper.
To reduce the latency and demand for storage space of the design, a magnitude-based post-training pruning strategy is employed here to reduce the number of the shift parameters\cite{8114708}.
It is to assume that the smaller parameters have less impact on the final output, i.e., less significant. Based on this assumption, the smaller parameters can be discarded without severe influence on the final performance of the model. 
The complete quantization process is briefly described in detail with pseudo code shown in Algorithm \ref{alg:shift}.
\begin{algorithm}
\caption{Shift Parameter Quantization}
\begin{algorithmic}[1]
  \Input
  \Desc{$Params$}{matrix of parameters}
  \Desc{$N$}{Number of parameters remained}
  \EndInput
  \Output
  \Desc{$Shifts$}{Quantized parameters}
  \Desc{$Signs$}{Sign of parameters}
  \EndOutput
  \\
    \State $Signs \gets \Call{sign}{Params}$
    \State $[rows, cols] \gets \Call{shape}{Params}$
    \For{$i = 1$ to $rows-1$}
        \For{$j = 1$ to $cols-1$}
            \State $ BitPositions \gets \Call{zeros}{N}$
            \State $ count \gets 0$
            \State $P_{binary} \gets \Call{binary}{AbsParams[i][j]}$
            \For{$i \gets 0$ \textbf{to} $\Call{len}{P_{binary}} - 1$}
                \If{$count < N$}
                    \If{$P_{binary}[i] = 1$}
                        \State $BitPositions[count] \gets i + 1$
                        \State $count \gets count + 1$
                    \Else
                        \State $BitPositions[count] \gets None$
                    \EndIf
                \EndIf
            \EndFor
            \State $Shifts[i][j][:] \gets BitPositions$
        \EndFor
    \EndFor
\end{algorithmic}
\label{alg:shift}
\end{algorithm}

Conventional quantization methods typically involve mapping weights to a limited integer range\cite{Gholami2021ASO}.
The output data of the quantized layer requires to be mapped back to the original value range before it can be used as input of the subsequent layers.
The proposed method does not require such operations. Thus, the hardware design does not need any DSP block to implement the floating-point multipliers.
As a result, the design can avoid potential throughput limitations related to DSP resources limitation.

\subsubsection{Encoding}
After quantization, the next step is to encode the parameters for storage.
In order to further compress the storage space required for parameters, offset binary encoding is used\cite{TODD201423}.
The offset binary encoding scheme introduces an offset to reposition the data within an optimal range before performing binary encoding.
In this scheme, the data is moved towards zero.
In the encoding process, we first extract and save the sign information of the parameters. 
Then, we perform quantization on the absolute value of the parameters. 
The bias is chosen based on the smallest non-zero number to ensure that it is moved to zero.
The parameters that are outside the data range of the majority of parameters are rounded to the nearest number. 
The pseudo code of the process is shown in Algorithm \ref{alg:binary}.

\begin{algorithm}
\caption{Biased Binary Encoding}
\begin{algorithmic}[1]
  \Input
  \Desc{$Params$}{matrix of parameters}
  \Desc{$N$}{Number of bits used for encoding}
  \EndInput
  \Output
  \Desc{$Encoded$}{Encoded parameters}
  \Desc{$Bias$}{Bias for the encoding}
  \Desc{$Signs$}{Sign of parameters}
  \EndOutput
  \\
    \State $MinValue \gets \Call{min}{Params}$
    \State $Bias \gets -MinValue $
    \State $MaxRange \gets 2^{N} - 1$
    \State $Signs \gets \Call{sign}{Params}$
    \State $AbsParams \gets \Call{abs}{Params}$
    \State $[rows, cols] \gets \Call{shape}{Params}$
    \For{$i = 1$ to $rows-1$}
        \For{$j = 1$ to $cols-1$}
            \State $P_{biased} \gets AbsParams[i][j] + bias$
            \If{$P_{biased} > MaxRange$}
                \State $P_{biased} \gets MaxRange$
            \EndIf
            \State $P_{encoded} \gets \Call{binary}{P_{biased}}$
            \State $Encoded[i][j] \gets P_{encoded}$
        \EndFor
    \EndFor
\end{algorithmic}
\label{alg:binary}
\end{algorithm}

\section{Experiment}
\begin{figure}[!t]
    \centering
    \subfloat{\includegraphics[width=2.5in]{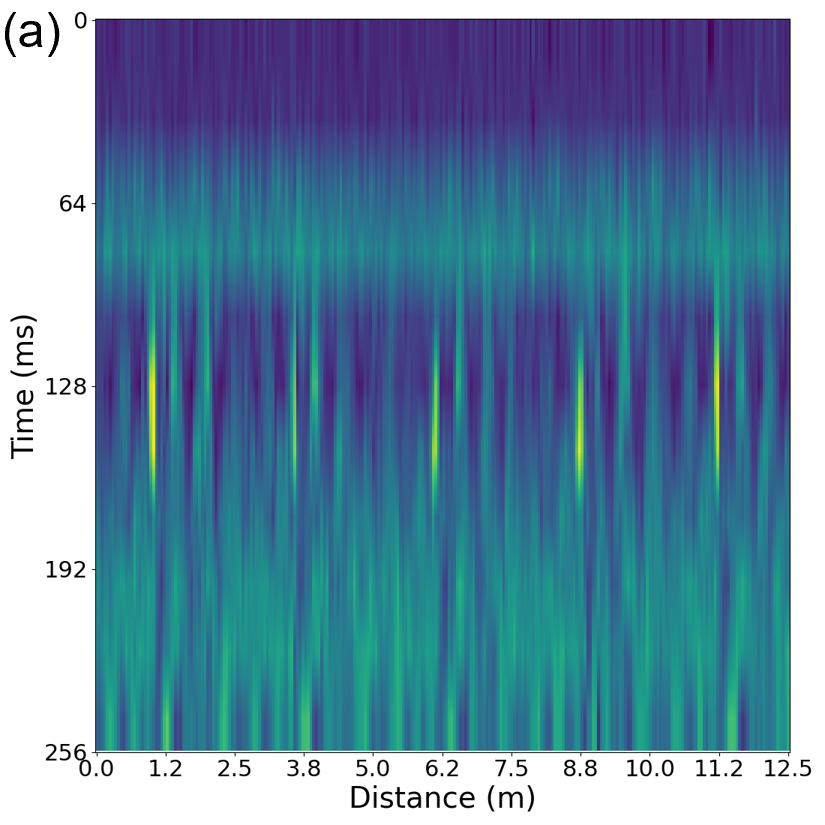}}\hfill
    \subfloat{\includegraphics[width=2.5in]{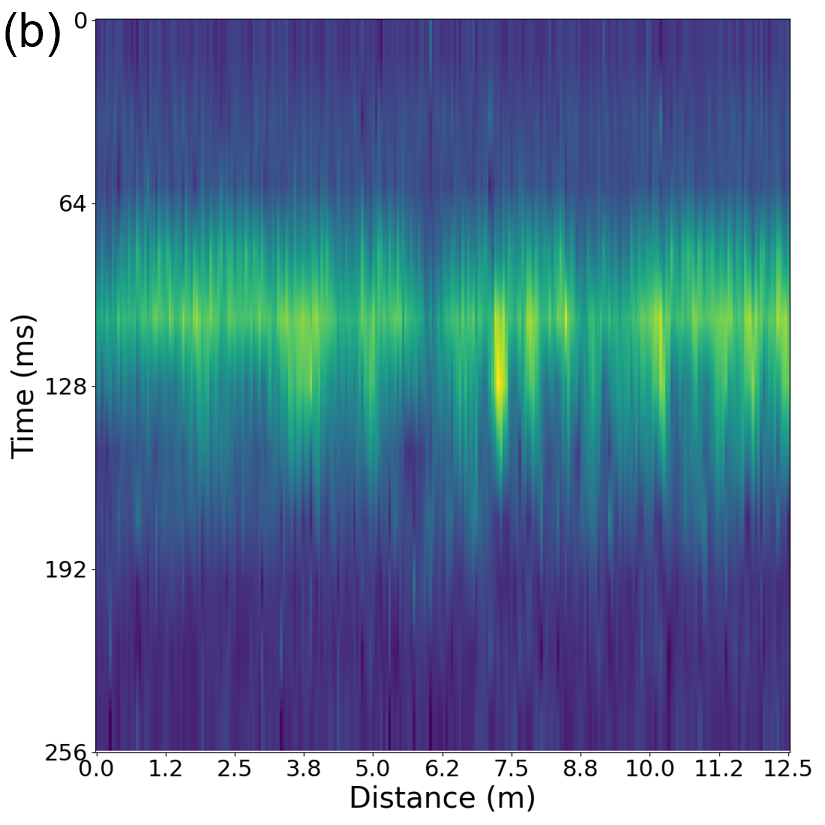}}\hfill
    \subfloat{\includegraphics[width=2.5in]{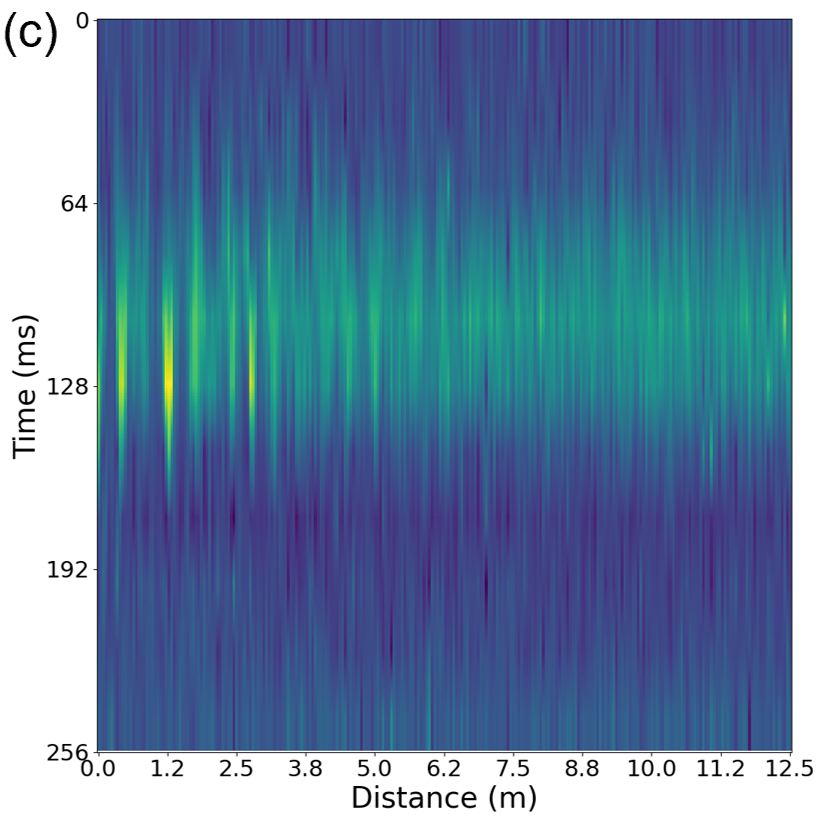}}
    \caption{Visualization of typical data samples. (a) Air Pick. (b) Excavator. (c) Hammer.}
    \label{fig:data_sample}
\end{figure}

The proposed scheme is evaluated using a real-world dataset captured by a DVS system, as illustrated in Fig. \ref{DVS_SYS}. The system capture a trace of Rayleigh back scattering light every 1 ms, with a spatial interval of 1.25 m. The samples stored as a 2-dimensional matrix with a size of 256 by 11, represent spatial-temporal figures, capturing a segment of the fiber with a length of 12.5 m and a time frame of 0.256 s. A 30 km long fiber was utilized, with a section of 50 m buried under 0.5 m of composite material composed of soil, sand, and stones in random ratios. The composition of the material has a significant impact on the vibration signal detected by the fiber, making it an effective means to explore the influence of environmental complexity. The dataset comprises two parts, collected from different locations at different time with varying material compositions, simulating the installation of the system in different environments for the same application. Specific details regarding the composition of the datasets are provided in Table \ref{dataset_tab}. Typical data are shown in Fig. \ref{fig:data_sample}.

\begin{table}
    \centering
    \caption{Class label and distribution of samples of the datasets}
    \begin{tabular}{|c|c|c|}
        \hline
        Class & Dataset 1 & Dataset 2\\
        \hline
        Hammer & 3332 & 268\\
        \hline
        Air Pick & 3558 & 330\\
        \hline
        Excavator & 3359 & 277\\
        \hline
    \end{tabular}
    \label{dataset_tab}
\end{table}

\subsection{Performance evaluation}
To evaluate and demonstrate the effectiveness of the proposed scheme, we employ a 5-fold cross-validation strategy on Dataset 1. This strategy involves dividing the dataset into five equally-sized subsets or folds. The division of folds remains consistent throughout the subsequent evaluations to ensure fair comparison of results. During each iteration of the cross-validation process, the model is trained on four folds while the remaining fold is held out for validation. This process is repeated five times, with each fold serving as the validation set once. The Dataset 2 is used as test set to verify the generalizability of the model. By averaging the performance of the five runs, we obtain the final validation and test result. Adam optimizer is used for training. Learning rate is set to 0.001, and it is halved when the training loss does not decrease within five epochs. The training setting is also kept consistent for the subsequent evaluations.

\begin{table}[http]
    \centering
    \caption{Parameters of the CNN Model}
    \begin{tabular}{|c|c|c|c|c|}
        \hline
        Layers & Kernel Size & Stride & Padding & Ouput Channels\\
        \hline
        Conv 1 & 3*3 & 1 & 1 & 8\\
        \hline
        MaxPool 1 & 2*2 & 2 & 0 & 8\\
        \hline
        Conv 2 & 3*3 & 1 & 1 & 16\\
        \hline
        MaxPool 2 & 2*2 & 2 & 0 & 16\\
        \hline
        Conv 3 & 3*3 & 1 & 1 & 32\\
        \hline
        AvgPool 1 & 2*2 & 2 & 0 & 32\\
        \hline
        Conv 4 & 3*3 & 1 & 1 & 64\\
        \hline
        flatten & - & - & - & 2048\\
        \hline
        FC 1 & - & - & - & 3\\
        \hline
    \end{tabular}
    \label{4lcnn}
\end{table}

To evaluate and demonstrate the performance of the lightweight model, a 4-layer CNN with a structure outlined in Table \ref{4lcnn} is utilized. The model consists of four layers and has a total of 30771 parameters. In terms of computational cost, it requires 2,282,496 floating-point operations (FLOPs). Additionally, this 4-layer CNN model serves as the baseline for evaluating the improvements in both performance and generalizability achieved through knowledge distillation.

\begin{figure}[!t]
     \centering
     \subfloat{     
         \centering
         \includegraphics[width=2.5in]{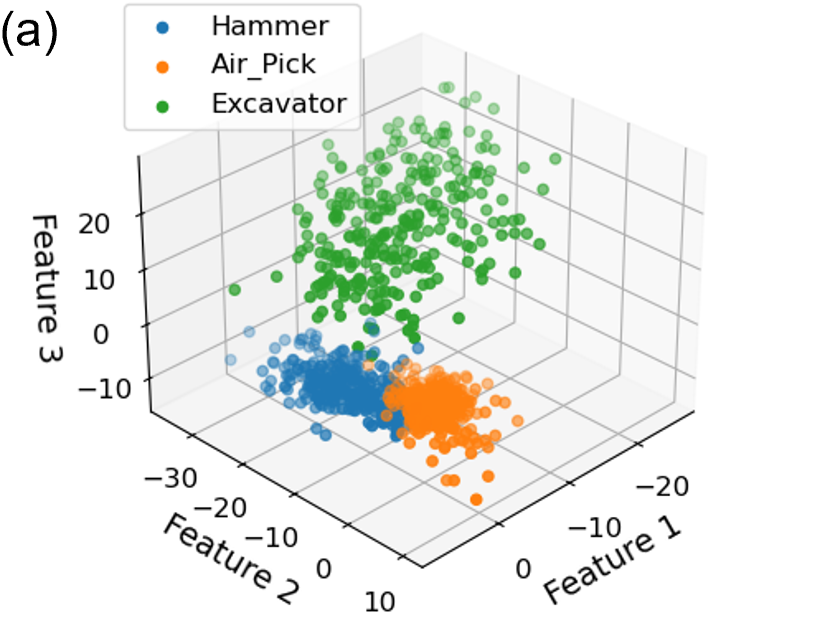}
         \label{fig:fv_1}
     }
     \hfill
     \subfloat{     
         \centering
         \includegraphics[width=2.5in]{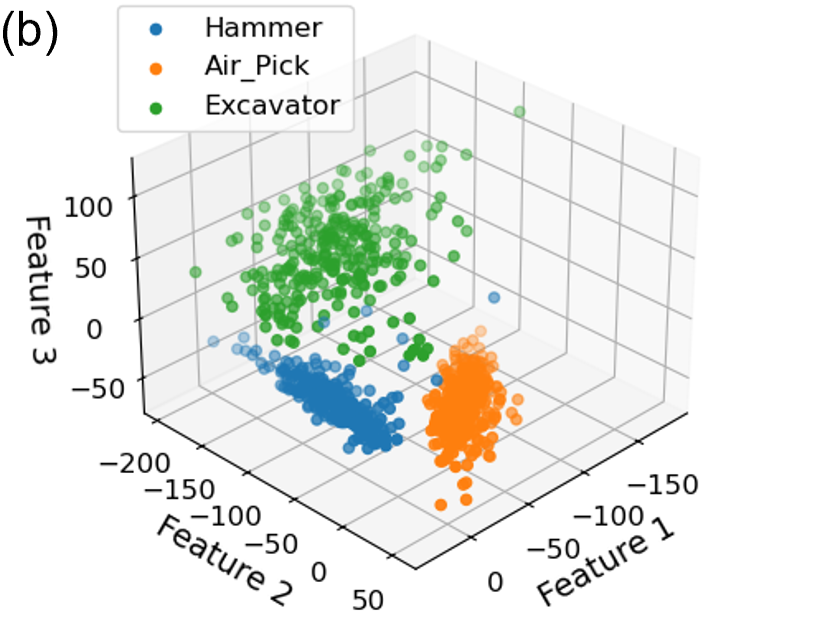}
         \label{fig:fv_2}
     }
     \hfill
        \caption{Feature visualization of the models. (a) Baseline Model. (b) Improved Model.}
        \label{fig:fv}
\end{figure}

The baseline model achieves an average validation accuracy of 99.69\%, while the average test accuracy is 83.41\%.
The feature visualization of the best model on the test set is shown in Fig. \ref{fig:fv_1}.  The drop in accuracy clearly demonstrates the limitation of generalizability of the baseline model.

\begin{table}
    \centering
    \caption{Evaluation Results of the Resnet Models}
    \begin{tabular}{|c|c|c|c|c|}
    \hline
        Model & Val/\% & Test/\% & FLOPs & Params\\
    \hline
        Resnet-18 & 99.58 & 95.47 & 98472960 & 2777283\\
    \hline
        Resnet-34 & 99.50 & 97.67 & 226235392 & 8164803\\
    \hline
        Resnet-101 & 98.34 & 85.87 & 857427968 & 27532227\\
    \hline
        Resnet-152 & 99.07 & 83.04 & 552325120 & 43175875\\
    \hline
    \end{tabular}
    \label{tab:res_result}
\end{table}

To ensure a strong baseline for knowledge distillation, we employed the Resnet model, which is widely used for this task in the relevant research\cite{Ge2022HighAccuracyEC, Jin2023PatternRO, Yao2023VibrationER}, as the teacher model. We further evaluated Resnet models with varying structures and depths on the dataset, and the results are presented in Table \ref{tab:res_result}. Considering our goal of improving both the performance and generalizability of the student model through knowledge distillation, we selected the Resnet-34 model as the teacher model. The knowledge distillation process introduces two additional hyper-parameters, namely $\alpha$ and $T$. To find the best combination, we iterate through different values of $\alpha$ from 0 to 1 with a step of 0.1, and $T$ from 1 to 10 with a step of 1. The optimal combination is found to be $\alpha = 0.1$ and $T = 5$.

\begin{figure}[h]
  \centering
  \includegraphics[width=2.5in]{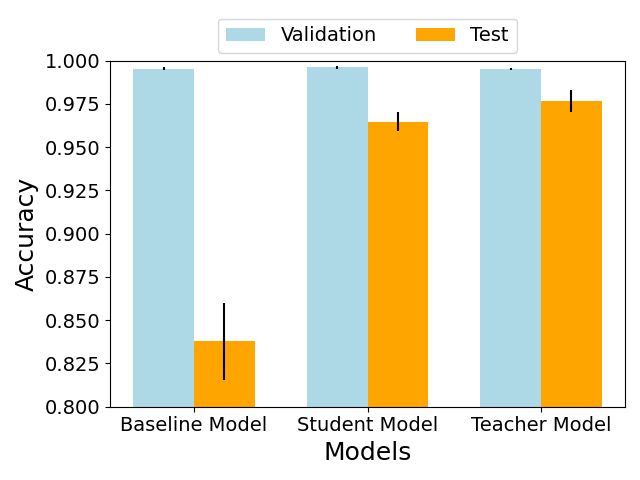}
  \caption{Validation accuracy and test accuracy of the models.}
  \label{fig:val_test}
\end{figure}

Through knowledge distillation, the model achieves an average test accuracy of 95.39\%, while maintaining an average validation accuracy of 99.61\%. The feature visualization on the test set is shown in Fig. \ref{fig:fv_2}, to better demonstrate the impact of knowledge distillation. It is clear that the model trained with knowledge distillation exhibits a more distinct boundary between samples belonging to different categories, demonstrating an improvement generalizability compared with the baseline model. The comparison of the baseline model, teacher model, and student model is presented in Fig. \ref{fig:val_test}. According to the figure, it is evident that the baseline model, which has fewer capacity and depth compared to the teacher model, exhibits limitations in terms of generalizability. The baseline model fails to extract high-level features that are less influenced by the training dataset, resulting in poor and unstable performance on the test dataset. It also shows that the knowledge distillation technique enables a lightweight CNN model, with only 30771 parameters and requiring 2282496 FLOPs, to achieve generalizability comparable to a larger model with 8164803 parameters and 226235392 FLOPs. 

\subsection{Implementation Evaluation}
\begin{figure}
     \centering
     \subfloat{     
         \centering
        \includegraphics[width=2.5in]{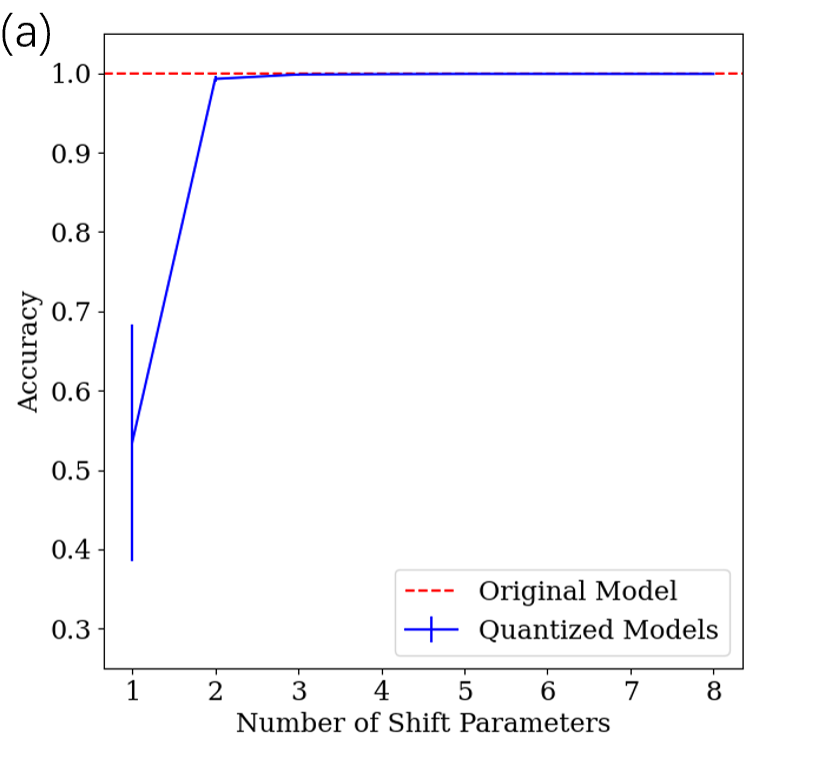}
        \label{fig:quan_val}
     }
     \hfill
     \subfloat{     
         \centering
        \includegraphics[width=2.5in]{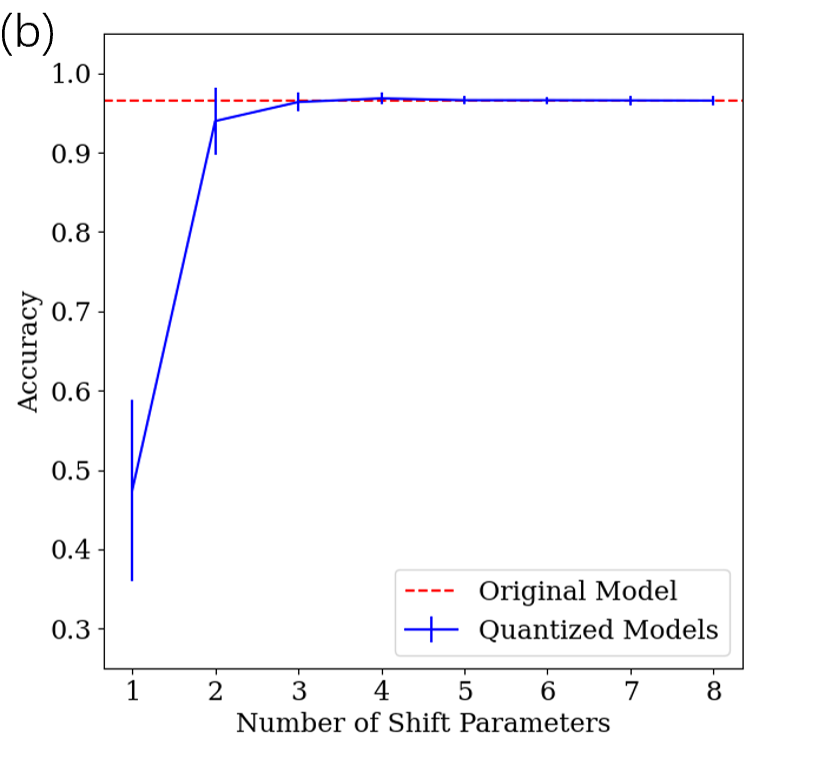}
        \label{fig:quan_test}
     }
     \hfill
        \caption{Change in accuracy with number of shift parameters: (a) Accuracy change on the validation set; (b) Accuracy change on the test set.}
        \label{fig:quan_both}
\end{figure}

We conduct a performance evaluation of the proposed quantization method on the resulting KD-improved models. The entire model is quantized together, meaning that all weights are quantized to the same number of shift parameters. During the quantization process, we iterate through different values of the number of shift parameters, ranging from 1 to 10. This enables us to assess and analyze the performance of the model across different degrees of quantization, thus allowing us to identify the optimal balance between compression level and performance degradation. The results of the quantization process are presented in Fig. \ref{fig:quan_val}, which shows the performance of the quantized models on the validation set. Based on the results, it can be determined that the optimal number of shift parameters to be used is 3. This allows the compressed model to achieve the same performance as the original model. To validate the generalizability of the model, the quantization process is further assessed on the test dataset, as demonstrated in Fig. \ref{fig:quan_test}. The results clearly indicate that the compressed model performs comparably to the original model, without any noticeable degradation in generalizability. This choice of quantization configuration effectively preserves all the essential information of the model while achieving the desired performance on the test dataset. 

\begin{figure}
     \centering
     \subfloat{     
         \centering
        \includegraphics[width=2.5in]{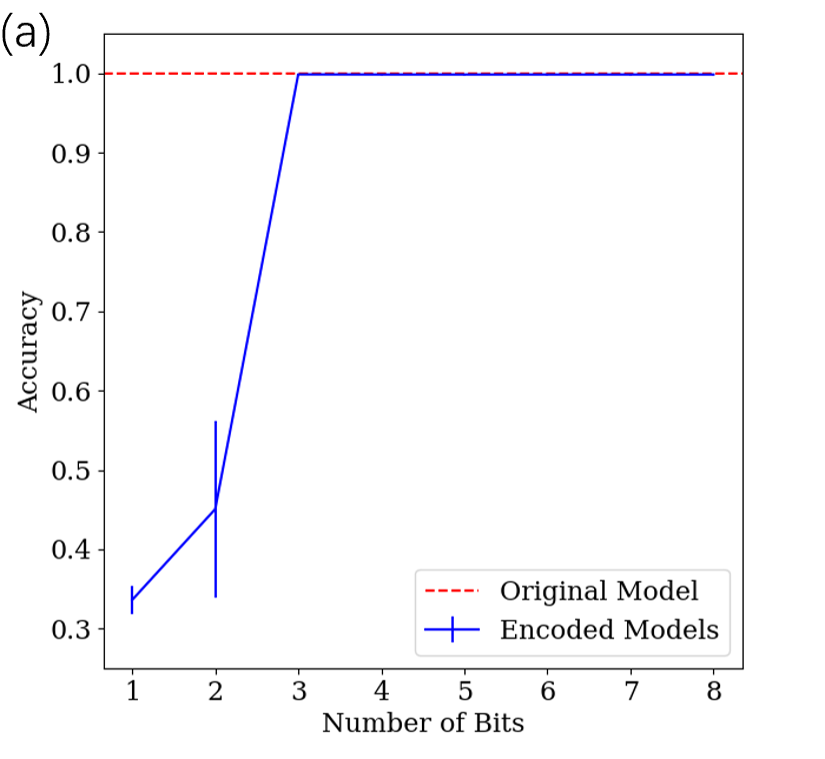}
        \label{fig:encode_val}
     }
     \hfill
     \subfloat{     
         \centering
        \includegraphics[width=2.5in]{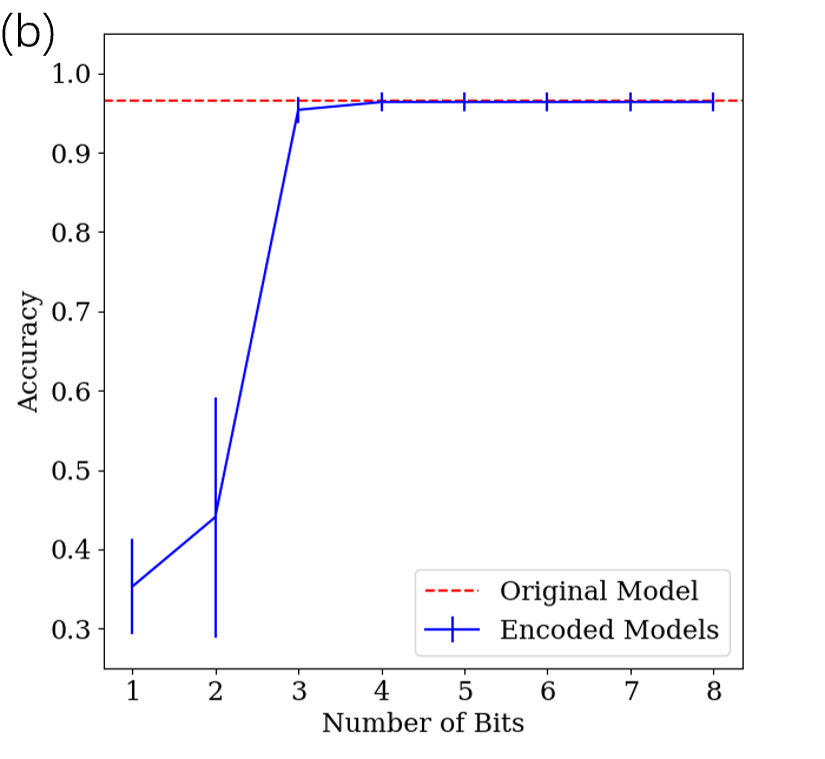}
        \label{fig:encode_test}
     }
     \hfill
        \caption{Change in accuracy with number of bits used for encoding: (a) Accuracy change on the validation set. (b) Accuracy change on the test set.}
        \label{fig:encode_both}
\end{figure}

The encoding process is then performed to optimize the on-chip storage. It is carried out on a layer-wise basis, which means that the parameters of the same layer are encoded together using the same bias. To determine the optimal number of bits used for encoding, we conduct iterations ranging from 1 to 8. The results obtained from the validation dataset, as depicted in Fig. \ref{fig:encode_val}, indicate that employing a 3-bit encoding scheme achieves lossless compression. Furthermore, the encoding process is performed on the test dataset and visualized in Fig. \ref{fig:encode_test}. It is evident that the encoding procedure brings a certain degree of performance degradation in terms of the generalizability of the model. Nevertheless, despite this influence, it is worth noting that the impact remains stable and minimal. Such stability in performance degradation implies that the encoding process does not significantly influence the overall performance of the model.

It is clear that the proposed approach offers an effective solution for achieving compression, with only a minimal impact on the generalizability of the model, as shown by the aforementioned results. We can then calculate the compression rate using the following information: the model initially consists of 30771 32-bit floating-point format parameters, and each weight is compressed into three 3-bit shift parameters. Consequently, the compression rate can be calculated: $3 \times 3/32 \times 100\% = 28.125\%$. Additionally, the proposed compression scheme has been adapted to the shift-add structure, further enhancing the overall system performance.

\begin{table}
    \centering
    \caption{Performance of the Models on Different Hardware}
    \begin{tabular}{|c|c|c|c|}
    \hline
        Model & Inference time & Inference time & Inference time\\
             & on CPU/ms      & on GPU/ms      & on FPGA/ms \\
    \hline
        ResNet-34 & 1.846 & 0.375 & -\\
        Quantized CNN & 0.254 & 0.178 & 0.083\\
    \hline
    \end{tabular}
    \label{tab:inf_time}
\end{table}

The optimized model is then implemented with proposed FPGA-based hardware acceleration scheme. To evaluate the proposed scheme, we use the state-of-the-art CPU and GPU as the baseline performance. Here, the utilized CPU is Intel i9-14900, GPU is Nvidia GTX 4090. The average inference time of them is shown in Table \ref{tab:inf_time}. The FPGA chip utilized here is Xilinx ZCU15eg. For quick development and fast implementation, the design is built with Xilinx high level HLS. The final design achieved a latency of 25112 cycles running at a frequency of 303 MHz. The inference time of the system can be calculated as $25112 \times 1/(303 \times 106) \approx 0.083 ms$. The sample time for each data frame is 256 ms, allowing the system to process 3,084 frames within this interval. This processing capacity is equivalent to handling the data generated when the system monitors a fiber with a length of $12.5 \times 3084 = 38550 m$. Therefore, the system can achieve real-time processing over a 38-km fiber, which is longer than the 30-km fiber utilized for sampling the dataset. Based on bibliographical research, the achieved recognition time of 0.083 ms is the fastest reported so far in the field \cite{yangRealTimeFOTDRVibration2022a, wuSmartFiberOpticDistributed2023}.

\begin{table}[htbp]
\caption{Hardware Resource Utilized}
\begin{center}
\begin{tabular}{|c|c|c|c|}
\hline
Resource & Utilization & Available\\
\hline
LUT & 280828 & 341280\\
FF & 257011 & 682560\\
DSP & 0 & 3528\\
BRAM & 11 & 744\\
URAM & 0 & 112\\
\hline
\end{tabular}
\label{tab_4}
\end{center}
\end{table}

The resource requirement of the design is shown in Table \ref{tab_4}. Our design makes efficient use of on-chip resources like look-up tables (LUT) and flip-flops (FF), allowing for high parallelism without being limited by the amount of DSP resources available. This optimization enables our system to achieve a higher performance compared with the CPU and GPU solutions.

\section{Conclusion}
This paper addresses the throughput limitations of DVS systems, which hinder the implementation of real-time recognition over long-distance fibers, which is essential for deploying DVS technology in smart IoT applications. To overcome this challenge, we propose using shallow CNN models as a more resource-efficient alternative to DLMs.
We identify the issue of limited generalizability in lightweight models and address it by applying knowledge distillation techniques to enhance their performance. Additionally, we propose a novel hardware acceleration scheme based on FPGAs to accelerate model inference.
Our approach achieves a significant reduction in inference time, which is 2.14 times faster than state-of-the-art GPU implementations and 4.52 times faster than CPU implementations, making it the fastest recognition method reported in this field to date. With this system, we enable real-time recognition over fibers up to 38 kilometers long, advancing the capabilities of DVS technology and its practical application as smart IoT systems.

\bibliographystyle{ieeetr}
\bibliography{jlt_2023_12}

\vspace{2ex}\noindent\textbf{Zhongyao Luo} received his BEng degree in Electronics and Computer Science from the University of Edinburgh in 2020. He is currently working toward a PhD at the School of Optical and Electronic Information, HUST, Wuhan, China. His current research interests include distributed optical fiber sensing, and machine learning.

\vspace{2ex}\noindent\textbf{Hao Wu} received his BS, MS, and PhD degrees from HUST, Wuhan, China, in 2013, 2016, and 2019, respectively. His postdoctoral research at HUST was focused on the machine earning algorithms for distributed optical fiber sensing. Since 2024, he has been a research associate at HUST. His current research interests are the integration of artificial intelligence and optical fiber.

\vspace{2ex}\noindent\textbf{Zhao Ge} received his BS degree from Jianghan University, Wuhan, China, in 2019 and his MD degree from HUST, Wuhan, China, in 2022. He is currently working toward a PhD at the School of Optical and Electronic Information, HUST, Wuhan, China.

\vspace{2ex}\noindent\textbf{Ming Tang} received his BEng degree from HUST in 2001 and his PhD from Nanyang Technological University, Singapore, in 2005. His postdoctoral research at the Network technology Research Center was focused on the optical fiber amplifier. From 2009, he was a research scientist in the Tera-Photonics Group, RIKEN, Japan. Since 2011, he has been a professor at HUST. His current research interests include optical fiber-based linear and nonlinear effects for communication and sensing applications.

\end{document}